\documentclass{article}
\usepackage{a4wide}
\usepackage{amssymb}
\usepackage{amsmath}
\usepackage{graphicx}

\newcommand{\iii}{\^\i}
\begin{document}
\begin{flushright}
{UB-ECM-PF-02-23
%MPI-PhT/2002-??,
%hep-ph/0210XXX }
}
\end{flushright}

\begin{center}
{\Large
\textbf{A modified Starobinsky's model of inflation:\\
Anomaly-induced inflation, SUSY and graceful exit.}}
\footnote{Talk given at SUSY 2002: the \textit{10th International
Conference on Supersymmetry and Unification of Fundamental
Interactions}, DESY, Hamburg, Germany, June 17-23; also at GRG 11:
\textit{International Conference on Theoretical and Experimental
Problems of General Relativity and Gravitation}, Tomsk, Russia,
July 1-8 2002. }

\vspace{0.8cm}

{\large {Ilya L. Shapiro}$^{a}$, \underline{Joan Sol\`a}$^{b,c}$}

\vspace*{0.2cm} {\sl $^a$ Departamento de F\'{\i}sica - Instituto
de Ciencias Exatas
\\
Universidade Federal de Juiz de Fora, 36036-330, MG, Brazil\\
$^b$  Departament d'Estructura i Constituents de la Mat\`eria,
Universitat de Barcelona, \\Diagonal 647, E-08028 Barcelona,
  Catalonia, Spain\\
$^c$ Institut de F\'{\i}sica d'Altes Energies, Universitat
Aut\`onoma de
  Barcelona, E-08193 Bellaterra, Barcelona, Catalonia, Spain}

\end{center}

\begin{center}
{ABSTRACT}
\end{center}
%\vskip 1mm

\begin{quotation}
\noindent {\sl We present a model of inflation based on the
anomaly-induced effective action of gravity in the presence of a
conformally invariant Hilbert-Einstein term. Our approach is
based on the conformal representation of the fields action and on
the integration of the corresponding conformal anomaly. In
contradistinction to the original Starobinsky's model, inflation
can be stable at the beginning and unstable at the end. The
instability is caused by a slowing down of inflation due to
quantum effects associated to the massive fermion fields. In
supersymmetric theories this mechanism can be linked to the
breaking of SUSY and suggests a natural way to achieve graceful
exit from the inflationary to the FLRW phase. } \vskip 1mm
\end{quotation}

\vskip 8mm
%\newpage

%%%%%%%%%%%%%%%%%%%%%%%%%%%%%%%%%%%%%%%%%%%%%%%%%%%%%%%%%%%%
%%%  \vskip 4mm
\noindent
{\large\bf Introduction}
\vskip 1mm

%\quad
Inflation \cite{Guth} automatically solves five of the six basic
cosmological problems\,\cite{{Peebles}}: 1) the monopole problem,
2) the horizon problem, 3) the flatness-curvature-entropy
problem, 4) the rotation problem, and 5) the large-scale
homogeneity versus small-scale inhomogeneity problem.  The
minimum number of e-folds of inflation required can be (roughly)
estimated in many different ways. As an example, take a flat
Friedmann-Lema\iii tre-Robertson-Walker (FLRW) model, then the
scale factor $a=a(t)$ in the matter-dominated (MD) and
radiation-dominated (RD) eras evolve as $ a\sim t^{2/3}$ and $
a\sim t^{1/2}$ respectively. Since at present  $t_0\sim 10^{18}$
sec ($15$ Gy) and $a(t_0)\sim 1.5\times 10^{10} lyr\sim 10^{28}cm
$, it follows that the scale factor at the end of the RD epoch
($t\sim 10^{12}$ sec $\sim 10^5$ yr) was
$a_R=a(t_0)/z_R=10^{24}cm$, where $z_R\sim a(t_0)/
a_R=\left(10^{18}/10^{12}\right)^{2/3}=10^4$ is the redshift at
that time. From this the scale factor at the Planck time ($\sim
10^{-44}$ sec) should be
$a^{*}_P=\left(10^{-44}/10^{12}\right)^{1/2}\,a_R\sim 10^{-4}cm$,
which is of course untenable! Therefore, to make this number to
match up the correct Planck length, $a_P\sim 10^{-33}$ cm, we need
to supplement the standard FLRW evolution with an early inflation
period in which the number of e-folds of inflation should be
around $\xi\sim 65$:
\begin{equation}\label{efold}
  e^{\xi}=\frac{a_P^*}{ a_P}=10^{29}\Rightarrow \xi\sim 67\,.
\end{equation}
At present, however, the classical cosmological problems that
motivated inflation are no longer regarded as the strongest
motivation for inflationary cosmology. For example, the relation
``homogeneity $\rightarrow$ flatness'' is not true. Nowadays we
have homogeneous open models of inflation. Besides providing the
solution to some cosmological problems, the important thing at
present is that inflationary models can be (and will be) more and
more accurately tested (something that one could not suspect 10
years ago) through their specific predictions on the metric and
density perturbations, which should be consistent with structure
formation and the anisotropies of the CMB\,\cite{CMBR}. These are
nowadays the real facts behind $\xi>65$ in Eq.(\ref{efold}),
rather than the previous and similar heuristic argumentations.

Inflation, however, does \textit{not} solve the sixth cosmological
problem, the cosmological constant (CC) problem\,\cite{weinRMP}.
The fact that the CC has been measured non-zero and
positive\,\cite{Supernovae} poses a new challenge and may require
novel approaches. One possibility is to think of the
renormalization group (RG) evolution of the cosmological
parameters \,\cite{JHEPCC1}. We will see that this same approach
can be applied to the study of cosmological inflation, if we
identify the RG scale with the expansion rate $\mu=H(t)$. This
identification can be understood as follows. At low energy the
dynamics of gravity is defined by Einstein's equations
\begin{equation}
R_{\mu \nu }-\frac{1}{2}\,Rg_{\mu \nu }=8\pi G_{N}\,(T_{\mu \nu
}+g_{\mu \nu }\,\Lambda\,)\,,  \label{Einstein}
\end{equation}
where $G_N=1/M_{P}^{2}$ is Newton's
constant. Let us use the value of the curvature scalar ($%
R$) to construct an order parameter for the gravitational energy.
By dimensional analysis the RG scale $\mu $ for gravity is
naturally associated with $R^{1/2}$. From Eq.(\ref{Einstein}) \
we see that this is equivalent to take $\mu \sim $ $\sqrt{T_{\mu
}^{\mu }/M_{P}^{2}}$. But in the cosmological setting the basic
dynamical equations refer to the scale factor $a(t)$ of the FLRW
metric, and so we must re-express the graviton energy in terms of
it. The $00$ component of (\ref{Einstein}) yields the well-known
Friedmann-Lema\iii tre equation
\begin{equation}
H^{2}\equiv \left( \frac{\dot{a}}{a}\right) ^{2}=\frac{8\pi }{3\,M_{P}^{2}}%
\left( \rho +\Lambda\right) -\frac{k}{a^{2}}\,.  \label{FL}
\end{equation}
The space curvature term can be safely set to zero ($k=0$). The
spatial components of (\ref{Einstein}), combined with the $00$
component (\ref{FL}), yields the following dynamical equation for
$a(t)$:
\begin{equation}
\ddot{a}=-\frac{4\pi }{3\,M_{P}^{2}}\left( \rho
+3\,p-2\,\Lambda\right) \,a\,.  \label{accel}
\end{equation}
In these equations $\rho =\rho _{M}+\rho _{R}$ is the total
energy density
of matter and radiation, and $p$ is the pressure. In the modern Universe $%
p\simeq 0$ and $\rho \simeq \rho _{M}^{0}$. Moreover, from the
recent supernovae data\thinspace \cite{Supernovae}, we know that
$\Lambda$ and $\rho _{M}^{0}$ have the same order of magnitude as
the critical density $\rho _{c}^{0}$. Therefore, the source term
on the \textit{r.h.s.} of (\ref {accel}) is characterized by a
single dimensional parameter $\sqrt{\rho _{c}^{0}/M_{P}^{2}}$,
which according to Eq. (\ref{FL}) is nothing but the
experimentally measurable Hubble's constant $H_{0}$. This is
obviously consistent with the expected result $\sqrt{T_{\mu
}^{\mu }/M_{P}^{2}}$ in the general case because $T_{\mu }^{\mu
}\sim \rho _{M}^{0}\sim \rho _{c}^{0} $ for the present-day
universe. \ Therefore, we conclude that the Ansatz
\begin{equation}
\mu \sim R^{1/2}\sim H(t)  \label{muH}
\end{equation}
is reasonable and we assume that this identification takes place
at each stage of the cosmological evolution. With this guiding
principle, a semi-classical description of gravitational
phenomena of quantum matter in a curved classical background
should be possible. In particular, in the early universe the
value of $H(t)$ decides which matter particles are active degrees
of freedom for the RG evolution of the parameters. Therefore,
particles whose masses satisfy $M>H(t)$ will be decoupled from
the relevant quantum effects at time $t$. Clearly, if one would
be able to concoct a natural mechanism by which $H(t)$
progressively slows down after the universe has achieved a
sufficient number of e-folds of inflation (namely $\xi>65$), then
inflation should eventually stop and the FLRW regime could
perhaps start. While many authors have looked for a suitable
scalar field (so-called ``inflaton'') capable of realizing this
scenario\cite{Guth}, in the original Starobinsky's
model\,\cite{star} inflation was attempted by looking for a
self-consistent solution of Einstein's equations when they are
modified to include the vacuum quantum effects. Unfortunately, in
Starobinsky's model inflation is unstable from the very beginning
(with the flat space stable), so that one has to fine-tune the
initial conditions to insure $\xi>65$ before inflation
stops\,\cite{vile}. It would be much desirable to find an improved
framework where the self-consistent solution appears first as a
stable inflationary solution (hence independent of the initial
conditions, even though space-time is unstable) and such that
subsequently (after $\xi>65$ is fulfilled) inflation becomes
unstable and the universe transits into the stable and flat FLRW
space-time. Indeed, a modified Starobinsky's model like that is
possible, provided that we can arrange the condition
$H(t)\rightarrow 0$ and at the same time distort the stability
regime thanks to a change in the number of active degrees of
freedom, e.g. due to a phase transition from a supersymmetric
Grand Unified Theory (GUT) into the Standard Model (SM) of the
strong and electroweak interactions. Such a scenario has been
first discussed in Ref.\cite{shocom}. It is based on a
modification of the anomaly-induced effective action\,\cite{star,
vile,hhr} resulting from a prior full conformization of the
classical action for gravitational fields\,\cite{deser70},
including the Hilbert-Einstein term, and matter
fields\,\cite{PSW}, and on the decoupling of the supersymmetric
particles at low energy\,\cite{insusy}. Furthermore, there are
strong indications that the spectrum and the amplitude of the
gravitational waves in this model \cite{wave,PST} are in
agreement with the existing CMBR data\,\cite{CMBR}. Recently,
also the stability with respect to small perturbations of the
conformal factor of the metric has been studied in the presence
of a cosmological constant\,\cite{JHEPCC1,PST}.

%%%%%%%%%%%%%%%%%%%%%%%%%%%%%%%%%%%%%%%%%%%%%%%%%%%%%%%%%%%%
%%%%%%%%%%%%%%%%%%%%%%%%%%%%%%%%%%%%%%%%%%%%%%%%%%%%%%%%%%%%
%%%%%%%%%%%%%%%%%%%%%%%%%%%%%%%%%%%%%%%%%%%%%%%%%%%%%%%%%%%%
\vskip 5mm \noindent {\large\bf 2. Local conformal invariance  and
effective action} \vskip 1mm

The expansion of an homogeneous, isotropic universe means a
conformal transformation of the metric
$\,g_{\mu\nu}(t)\rightarrow a^2(\eta)\,{g}_{\mu\nu},$ where
$a(\eta)=\exp\,\sigma(\eta)$ and $\eta$ is the conformal time
($d\eta = dt/a(\eta)$).  Suppose that one starts from the
conformal invariant formulation of the Standard Model of the
strong and electroweak interactions \cite{PSW,Coleman} and of
gravity\cite{deser70}, and then one uses the well-known methods to
derive the anomaly-induced action\,\cite{reigert,book}. It is
natural to think that the latter can be applied at high energies,
where the masses of the matter fields are negligible. At the
classical level, the theory which results from this procedure is
always equivalent to the original theory. Nevertheless, in the
quantum theory the equivalence is destroyed by the anomaly, which
can be calculated explicitly. In particular, the massive fields
may also contribute to it. Besides the anomalous terms, there are
the conformal invariant quantum corrections to the classical
vacuum action. Our first purpose is to construct such a
formulation of the SM in curved space-time which possesses local
conformal invariance in $d=4$. Actually, the procedure can be
applied to any gauge theory, e.g. the SM and extensions thereof,
including GUT's and supersymmetric generalizations like the
Minimal Supersymmetric Standard Model (MSSM)\,\cite{MSSM}.

The original action of the theory includes kinetic terms for
spinor and gauge boson fields, as well as interaction terms, all
of them already conformal invariant. As for scalars (e.g. Higgs
bosons) we suppose that their kinetic terms appear in the
combination $\,g^{\mu\nu}\partial_\mu\varphi
\partial_\nu\varphi + 1/6\cdot R\varphi^2\,$ providing the local
conformal invariance. The non-invariant terms are the massive
ones for the scalar and spinor fields, but also the
Hilbert-Einstein term giving General Relativity at low energies.
In all these cases the conformal non-invariance is caused by the
presence of dimensional parameters $\,\,m_H^2$, $\,m$,
$\,M^2_P=1/G$. The central idea is to replace these parameters by
functions of some new auxiliary scalar field $\,\chi$. For
instance, we replace\,\cite{PSW}
\begin{eqnarray}
m_H^2 \to \frac{m_H^2}{M^2}\,\chi^2 \,,\,\,\,\,\,\,\,\, m \to
\frac{m}{M}\,\chi \,,\,\,\,\,\,\,\,\, M_P^2 \to
\frac{M_P^2}{M^2}\,\chi^2\,, \label{replace}
\end{eqnarray}
%$m_H^2 \to (m_H^2/{M^2})\,\chi^2$, $m \to ({m}/{M})\,\chi$ and
%$M_P^2 \to ({M_P^2}/{M^2})\,\chi^2$
where $M$ is some dimensional
parameter, e.g. related to a high scale of spontaneous breaking of
dilatation symmetry\,\cite{PSW}. Then the scalar and fermion mass
terms become quartic interactions and Yukawa couplings
respectively,
\begin{eqnarray}
\frac12\int
d^4x\sqrt{-g}\,\,\frac{m_H^2}{M^2}\,\varphi^2\,\chi^2\,,
\,\,\,\,\,\,\,\,\,\,\,\,\,\,\,\,\,\,\,\,\, \int
d^4x\sqrt{-g}\,\,\frac{m}{M}\,\,{\bar\psi}\psi\,\chi\,,
\label{masses}
\end{eqnarray}
which are of course (local and global) conformal invariant.
Furthermore, the Hilbert-Einstein term gets conformized too:
\begin{eqnarray}
S^*_{EH} = -\frac{1}{16\pi G\,M^2}\,\int d^4 x\sqrt{-g}\,
\left[\,R\chi^2 + 6\,(\partial \chi)^2\,\right]\,. \label{new
gravity}
\end{eqnarray}
After setting $\,\chi \to M\,$ this expression becomes identical
to the ordinary gravitational term, and from (\ref{masses}) the
ordinary mass terms for scalars and fermions are recovered at the
same time . This fixing can be called ``conformal unitary gauge''
in analogy with the unitary gauge of ordinary gauge theories, and
the scale $M$ can be associated to the vacuum expectation value
of the spontaneously broken dilatation symmetry at high energies
\cite{PSW}

It is supposed that the new scalar field $\,\chi\,$ takes the
values close to $M$, especially at low energies. But, there is a
great difference between $\,\chi\,$ and  $\,M\,$ with respect to
the conformal transformation. The mass does not transform, while
$\,\chi\,$ does. Then, the action of the new model becomes
invariant under the conformal transformation
\begin{eqnarray}
\chi \to \chi\,e^{-\sigma}\,,
\label{auxiliar}
\end{eqnarray}
which is
performed together with the usual transformations for the
other fields
\begin{eqnarray}
g_{\mu\nu}\to
g_{\mu\nu}\,e^{2\sigma} \,,\,\,\, \varphi \to \varphi\,e^{-\sigma}
\,,\,\,\, \psi \to \psi\,e^{-3/2\,\sigma} \,.
\label{conformal}
\end{eqnarray}
Thus, in the matter sector our program of ``conformization'' is
complete. When we quantize the theory, it is important to
separate the quantum fields from the ones which represent a
classical background. In order to maintain the correspondence
with the usual formulation of the SM, we avoid the quantization
of the field $\chi$ which will be considered, along with the
metric, as an external classical background for the quantum
matter fields. It is well known (see, e.g. \cite{book}) that the
renormalizability of the quantum field theory in external fields
requires some extra terms in the classical action of the theory.
The list of such terms includes the nonminimal term of the
$\,\int R\varphi^2$-type in the Higgs sector, and the action of
external fields with the proper dimension and symmetries. The
higher derivative part of the vacuum action has the form
\begin{eqnarray}
S_{vac} = \int d^4 x\sqrt{- g}\, \left\{l_1C^2 + l_2E +
l_3{\nabla^2}R\,\right\}\,,
\label{vacuum}
\end{eqnarray}
where, $l_{1,2,3}$ are some parameters, $C^2$ is the square of
the Weyl tensor and $E$ is the integrand of the Gauss-Bonnet
topological invariant. Now, since there is an extra field
$\,\chi$, the vacuum action should be supplemented by the
$\,\chi$-dependent term. The only possible, conformal and
diffeomorphism invariant, terms with dimension $4$ are (\ref{new
gravity}) and  the$\,\int\chi^4$-term. The last contributes to
the cosmological constant, which we suppose to cancel and do not
consider here. Its effect is reported
elsewhere\,\cite{JHEPCC1,PST}.

The next step is to derive the conformal anomaly in the theory
with two background fields $\,g_{\mu\nu}\,$ and  $\,\chi\,$. The
anomaly results from the renormalization of the vacuum action
including the terms (\ref{new gravity}) and (\ref{vacuum}). For
the sake of generality, let us suppose that there is also some
background gauge field with strength tensor $\,F_{\mu\nu}$. Then
the conformal anomaly has the form
\begin{eqnarray}
<T_\mu^\mu> \,=\, - \,\Big\{
\,wC^2 + bE + c{\nabla^2} R+d F^2
+ f\,[\,R\chi^2 + 6\,(\partial \chi)^2\,] \Big\}\,,
\label{anomaly}
\end{eqnarray}
where $\,w,\,b,\,c\,$ are $\,\beta$-functions for the parameters
$\,l_1,\,l_2,\,l_3$; $\,\,f\,\,$ is the $\beta$-function for the
dimensionless parameter  $\,\,1/(16\pi G\,M^2)$ of the action
(\ref{new gravity}), and  $\,d\,$ is the $\beta$-function for the
gauge coupling constant. The values of $\,\,w,b\,$ and $\,c\,$
depend on the matter content ($N_i$ being the number of particles
with spin $i$):
\begin{equation}
w=\frac{N_0 + 6N_{1/2} + 12N_1}{120\cdot (4\pi)^2 }\,\,,
\,\,\,\,\,\,\, b= -\,\frac{N_0 + 11N_{1/2} + 62N_1}{360\cdot
(4\pi)^2 }\,\,, \,\,\,\,\,\,\, c=\frac{N_0 + 6N_{1/2} -
18N_1}{180\cdot (4\pi)^2}\,. \label{abc}
\end{equation}
Recall that the condition for stable inflation is
$c>0$\,\cite{star}. Then one can play with various models. For
instance, from the previous equation it follows that the particle
content  of the SM ($N_0=4,N_{1/2}=24,N_1=12$) leads to $c<0$
(unstable inflation), which suggests that with only SM matter
fields the inflation period cannot be easily sustained, and it
might be insufficient\,\cite{star}.  On the other hand, for the
MSSM\,\cite{MSSM} ($N_0=104,N_{1/2}=32,N_1=12$) one has $c>0$
(stable inflation) etc. Clearly, we need physics beyond the SM in
order to possibly arrange a graceful transit from a regime of
stability (insuring the condition $\xi>65$) into another of
instability that might hopefully end into the FLRW phase.

We will argue that the presence of the non-zero $\beta$-function
$f$ could be the necessary quantum dynamical mechanism for the
graceful exit. From direct calculation using the Schwinger-DeWitt
method (see e.g. \cite{book}) we get
\begin{eqnarray}
f\,=\,\sum_{i}\,\frac{N_i}{3\,(4\pi)^2}\,\frac{m_i^2}{M^2}\,,
\label{f}
\end{eqnarray}
where $N_i$ are the number of Dirac spinors with
masses $m_i$. We note that bosons do not contribute to $f$.

In order to obtain the anomaly-induced effective action, we put
$\,{g}_{\mu\nu} = {\bar g}_{\mu\nu}\cdot e^{2\sigma}\,$ and
$\,\chi = {\bar \chi}\,\cdot e^{-\sigma}\,$, where the metric
$\,{\bar g}_{\mu\nu}\,$ has fixed determinant and the field
$\,{\bar \chi}\,$ does not change under the conformal
transformation. Then, the solution of the equation for the
effective action $\bar \Gamma$ proceeds in the usual way
\cite{reigert}. Disregarding the conformal invariant term we
arrive at the following expression\,\cite{shocom}:
$$
{\bar \Gamma} = \int d^4 x\sqrt{-{\bar g}} \,\{w{\bar C}^2 + b
({\bar E}-\frac23 {\bar \nabla}^2 {\bar R})
+ 2 b\,\sigma{\bar \Delta} + d{\bar F}^2 +
$$
\begin{eqnarray}
+ f\,[\,{\bar R}{\bar \chi}^2 + 6\,(\partial {\bar
\chi})^2\,]\,\} \sigma - \frac{3c+2b}{36}\,\int d^4
x\sqrt{-g}\,R^2\,. \label{quantum}
\end{eqnarray}

%%%%%%%%%%%%%%%%%%%%%%%%%%%%%%%%%%%%%%%%%%%%%%%%%%%%%%%%%%%%
\vskip 6mm \noindent {\large\bf 3. The role of masses in slowing
down inflation} \vskip 2mm
 In order to understand the role
of the particle masses in the anomaly-induced inflation, let us
consider the total action with quantum corrections
\begin{eqnarray}
S_t = S_{matter} + S_{EH} + S_{vac} + {\bar \Gamma}\,.
\label{totality}
\end{eqnarray}
One of the approximations we made was to disregard higher loop
and non-perturbative effects in the vacuum sector.  Another
approximation is that we take only the leading-log corrections.
Usually, this is justified if the process goes at high energy
scale. If the quantum theory has UV asymptotic freedom, the higher
loops effects are suppressed, and our approximation is reliable.
At the low-energy limit, we suppose that the massive fields
decouple and  their contributions are not important. Then Eq.
(\ref{totality}) can be presented in the form
\begin{eqnarray}
&& S_t=\int d^4 x\sqrt{-{\bar g}}\,\Big\{\,
\Big(\,-\frac{M_P^2}{16\pi M^2} + f\sigma\,\Big) \,[\,{\bar R}{\bar
\chi}^2 + 6\,(\partial {\bar \chi})^2\,]\nonumber\\ && -
\Big(\,\frac{1}{4} - d\sigma\,\Big)\,{\bar F}^2\, \Big\} +
S_{matter} +\,high.\, deriv.\, terms\,.
\label{label}
\end{eqnarray}
One can see that the modifications with respect to the case of
free massless fields \cite{wave} are an additional $f$-term and
the contribution to anomaly due to the background gauge fields.
 In order to restore the
Hilbert-Einstein term and get the inflationary solution, we fix
the conformal unitary gauge and put $\,\chi = {\bar
\chi}\,e^\sigma =M$. Furthermore, we can choose the conformally
flat metric ${\bar g}_{\mu\nu} = \eta_{\mu\nu}$. Then the
gravitational part of the action (\ref{label}) becomes
$$
S_{grav} = \int d^4x\,\Big\{\,2b\,(\partial^2\sigma)^2
- (3c+2b)\,[(\partial\sigma)^2+ \partial^2\sigma)]^2 -
$$
\begin{eqnarray}
- \,6M_P^2\,e^{2\sigma}\,(\partial\sigma)^2\,
\Big[\,1-\frac{16\pi M^2}{M^2_P}\,f\,\Big] - \Big(\,\frac{1}{4} -
d\sigma\,\Big)\,{\bar F}^2\, \Big\} \,.
\label{flat}
\end{eqnarray}
Computing the equation of motion of $a=\ln\sigma$ in terms of the
physical time $t$ (where $dt = a(\eta)d\eta$) we find
$$
a^2{\stackrel{....} {a}}
+3a{\stackrel{.} {a}}{\stackrel{...} {a}}
- \left(5 + \frac{4b}{c}\right){\stackrel{.} {a}}^2{\stackrel{..}{a}}
+ a{\stackrel{..} {a}}^2
- \frac{M_{P}^{2}}{8\pi c}\left( a^2 {\stackrel{..} {a}}
+ a{\stackrel{.} {a}}^2\right)+
$$
\begin{eqnarray}
+\frac{2fM^2}{c}\ln a \left( a^2 {\stackrel{..} {a}} +
a{\stackrel{.} {a}}^2\right)+\frac{2fM^2}{c}\,\frac{{\dot a}^2}{a}
- \frac{d{\bar F}^2}{6ca}=0.
\label{for t}
\end{eqnarray}
An exact solution of (\ref{for t}) does not look possible, but it
can be easily analyzed within the approximation that $f$ is not
too large. Then the new terms (collected in the second line of
Eq. (\ref{for t})) can be considered as perturbations. Moreover,
the last two of them are irrelevant, because during inflation
they decrease exponentially with respect to the other terms.
Thus, in this approximation, the only one relevant change is the
replacement $ M_P^2 \longrightarrow M_P^2\left[1- \tilde{f}\,\ln
a(t)\right]$ where for future convenience we have introduced the
dimensionless parameter
\begin{eqnarray}
\tilde{f}\equiv \frac{16\pi f\,M^2}{M_P^2}= \sum_i\,\frac{N_i}{3
\pi}\,\frac{m_i^2}{M_P^2}\,. \label{ftilde}
\end{eqnarray}
Notice that $f$ is given by Eq. (\ref{f}) and so $\tilde{f}$ does
not depend on the scale $M$. Since $f$ is small, the effect of
the masses may be approximated through the modification of the
inflation law $ a(t) = a_0\,e^{H_1t} $ according to\,\footnote{We
remind the reader that the coefficient $\,b\,$ is negative for
any particle content, see (\ref{abc}).}
\begin{eqnarray}
H_1 = \frac{M_P}{\sqrt{- 16\pi b}} \,\,\longrightarrow \,\,
\frac{M_P}{\sqrt{- 16\pi b}}\left[ 1- \tilde{f}\,\ln
a(t)\right]^{1/2}
%%% \sqrt{\,\frac{M_P^2}{-16\pi b} - M^2\,f\,\ln a(t) }
= H(t)\,,
\label{hawk}
\end{eqnarray}
To substantiate our claim, we have solved Eq. (\ref{for t})
directly using the numerical methods. The plots corresponding to
the numerical solution of the Eq. (\ref{for t}) are shown in
Fig.\,1. Since in the first period of inflation masses do not
play much role and the stabilization of the exponential inflation
proceeds very fast, the initial data (in both Eq. (\ref{hawk})
and the plots of Fig.\,1) were chosen according to the
exponential inflation law. According to the numerical analysis,
the total number of $e$-folds in the ``fast phase'' of inflation
(until the Hubble constant becomes comparable to the transition
scale $M^{*}$ where instability develops) is about $10^4$ for our
particular values of the parameters, and at the last stage the
expansion essentially slows down.
%%%%%%%%%%%%%%%%%%%%%%%%%%%%%%%%%%%%%%%%%%%%%%%%%%%%%%%%%%%%%%%%%%%%%%%%%%%%%%%%%%%%%%%
\begin{center}
\begin{figure}[tb]
\begin{tabular}{cc}
\mbox{\hspace{1.5cm}} (a) & \mbox{\hspace{0.5cm}} (b) \\
\mbox{\hspace{1.0cm}}\resizebox{!}{4cm}{\includegraphics{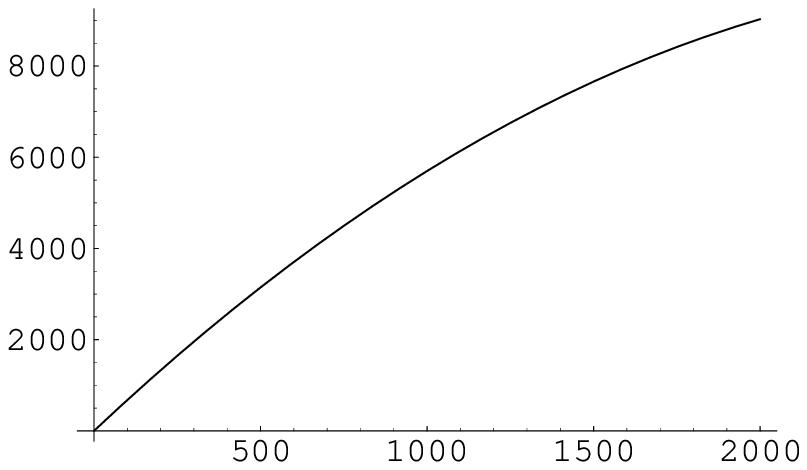}}
& \resizebox{!}{4cm}{\includegraphics{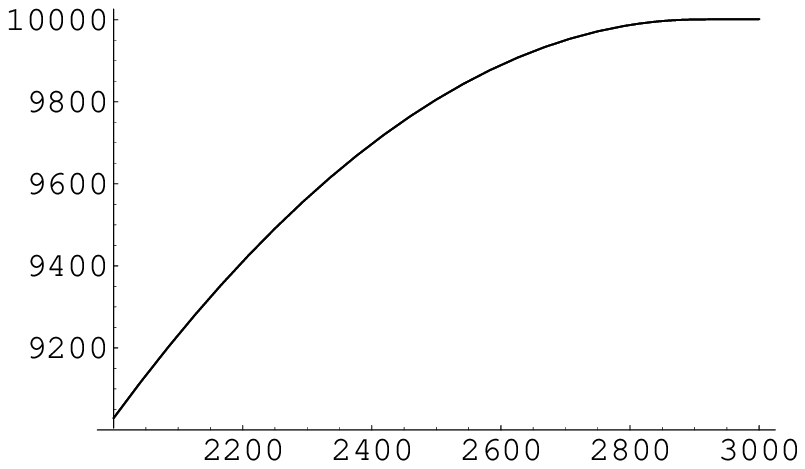}}
\end{tabular}
\begin{quotation}
\noindent \textbf{Figure 1}.\textsl{ \textbf{(a)} Plot of\,
$\sigma=\ln(a)$ versus the physical time $t$ as a result of the
numerical analysis of Eq.(\ref{for t}); $t$ is given in units of
$16\,\pi/M_P$ and we fixed the parameter (\ref{ftilde}) as
$\tilde{f}=10^{-4}$. Initial data: $a(0)=1\,,{\stackrel{.} {a}}(0)
= H_1 \,,{\stackrel{..} {a}}(0) = H_1^2\,,{\stackrel{...} {a}}(0)
= H_1^3\,.$
 In this time interval, inflation does not stop, yet; \textbf{(b)} As in
(a), but extending the numerical analysis until reaching an approximate
plateau marking the end of stable inflation.}
 \end{quotation}
 \label{fig1}
\end{figure}
\end{center}
%%%%%%%%%%%%%%%%%%%%%%%%%%%%%%%%%%%%%%%%%%%%%%%%%%%%%%%%%%%%%%%%%%%%%%%%%%%%%%%%%%%%%%%%%
%%%%%%%%%%%%%%%%%%%%%%%%%%%%%%%%%%%%%%%%%%%%%%%%%%%%%%%%%%%%%%%%%%%%%%%%%%%%%%%%%%%%%%%%
\begin{center}
\begin{figure}[tb]
\begin{tabular}{cc}
\mbox{\hspace{1.5cm}} (a) & \mbox{\hspace{0.5cm}} (b) \\
\mbox{\hspace{1.0cm}}\resizebox{!}{4cm}{\includegraphics{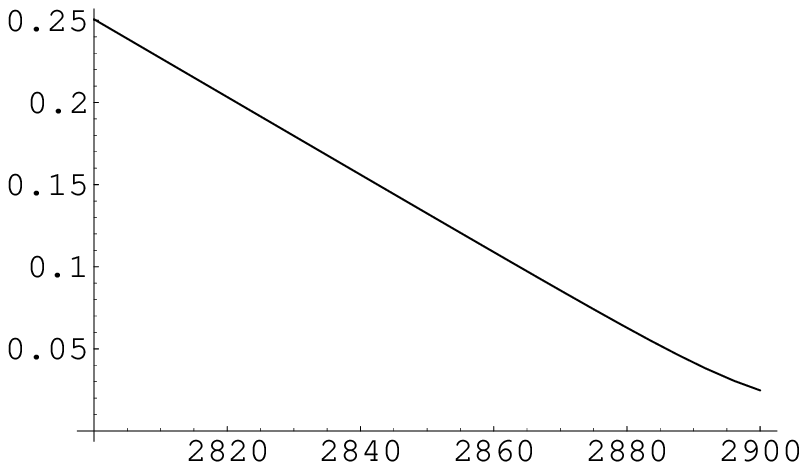}}
& \resizebox{!}{4cm}{\includegraphics{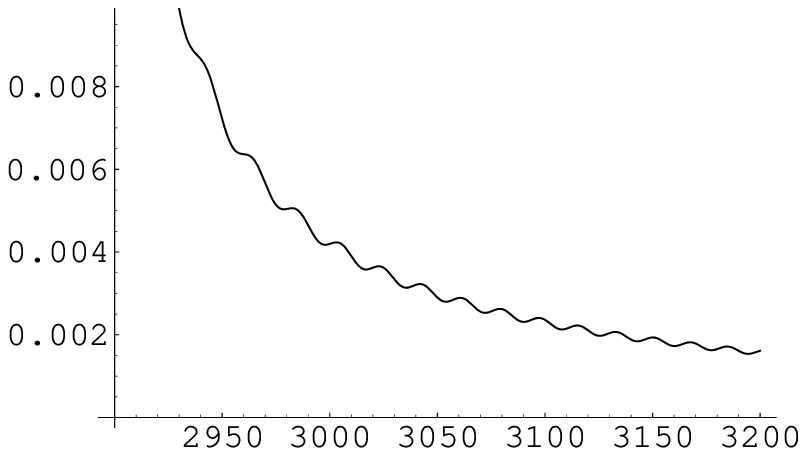}}
\end{tabular}
\begin{quotation}
\noindent \textbf{Figure 2}. \textsl{ Plot of
$H(t)=\dot{a}(t)/a(t)$ versus $t$ as a result of the numerical
analysis of Eq.(\ref{for t}) and parameter values as in Fig.1:
\textbf{(a)} $H(t)$ near the onset of the plateau; \textbf{(b)}
$H(t)$ well over the plateau.}
 \end{quotation}
 \label{fig2}
\end{figure}
\end{center}
%%%%%%%%%%%%%%%%%%%%%%%%%%%%%%%%%%%%%%%%%%%%%%%%%%%%%%%%%%%%%%%%%%%%%%%%%%%%%%%%%%%%%%%%%
 The chosen value of the parameter
(\ref{ftilde}) $\tilde{f}=10^{-4}$ in the plot is, as we warned
before, independent of the scale $M$, and it determines where the
process of stable inflation finishes as well as the number of
$e$-folds. On the other hand, if the scale $M^{*}$ is chosen near
the typical SUSY GUT value $M_{SUSY}\sim M_X \approx
10^{16}\,GeV$ for supersymmetry breaking at high energies, some of
the spinor masses $m_i$ will be of order $M_{SUSY}\approx
10^{16}\,GeV$. The latter assumption is indeed sound because the
$m_i$ will include the supersymmetric fermions associated to the
super-heavy gauge and Higgs bosons at the GUT scale, and so
$\,m_i\sim M_X\sim M_{SUSY}$.   From Eq. (\ref{ftilde}) it follows
that the parameter $\tilde{f}$ will be numerically smaller than
the one we have assumed in Fig. 1, and consequently the amount of
inflation will be larger. But the important qualitative point is
that for any value of $\tilde{f}$ the approximate plateau
eventually appears and signals the end of stable inflation. Also
notice from Fig.\,1 that the initial evolution is close to the
exponential inflation, but after that the expansion slows down
due to the quantum effects of massive fermions.
%%%%%%%%%%%%%%%%%%%%%%%%%%%%%%%%%%%%%%%%%%%%%%%%%%%%%%%%%%%%
\vskip 6mm
\noindent
{\large\bf 4. Graceful exit from anomaly-induced inflation}
\vskip 2mm

Recall from Eq.(\ref{muH}) that $\,H(t)$ sets the scale of the RG
running for the gravitational part. So if we consider the SUSY
breaking and the corresponding change in the number of active
degrees of freedom, then the necessary and sufficient condition
for the applicability of our approach is that $H(t)$ decreases
from the initial value about\,\footnote{Notice that $\,|16\pi b|
= {\cal O}(1)\,$ in the MSSM, and it is much larger than $1$ in
any typical SUSY GUT.} $M_P/\sqrt{-16\pi b}\sim 10^{18}\,GeV$,
down to the lower scale $\,H=M^{*}\lesssim M_{SUSY}$. The outcome
is that the evolution according to (\ref{hawk}) lasts until
reaching the scale $M^{*}$, and after that most of the SUSY
particles are decoupled, the inflationary solution becomes
unstable and the FLRW phase can start. In fact, the crucial point
is the existence of a nonvanishing $f$ as it eventually tempers
stable inflation allowing favorable conditions for the universe
to tilt into the FLRW phase.

To justify our claim that $M^{*}<M_{SUSY}$, recall that for a
really successful exit from the inflationary phase we need to make
sure that the amplitude of the gravitational waves is consistent
with the observable range of anisotropy in the CMBR. This will be
the case if during the last $65$ $e$-folds of the inflation, the
expansion rate $H(t)$ does not exceed $10^{-5}M_P$. Then the
fluctuations in the amplitude $h$ of these waves, $\delta
h/h=H/M_P$, will preserve the measured fluctuations in the
temperature of the relic radiation according to the relation
$\delta h/h=\delta T/T={\cal O}(10^{-5})$. At the lowest end of
the inflation interval this condition corresponds, in our
framework, to fix the instability value $M_{*}\approx 10^{-5}
M_P=10^{14} GeV$. It means that, in reality, we expect that after
the onset of the approximate plateau in Fig. 1 (b), where SUSY
breaking occurs, the universe will take a while before entering
the FLRW phase, i.e. the latter will actually initiate at some
point well over the plateau where $H=M_{*}\sim 10^{-5} M_P $. To
better assess this issue we have numerically analyzed $H(t)$ over
the plateau, see Fig. 2. We see that $H(t)$ decreases very fast on
it. For instance, from the comparison of Fig. 2(a) and Fig. 2(b),
we find that a $15\%$ increase of the time after the onset of the
plateau amounts $H(t)$ to diminish two orders of magnitude. So in
general $H(t)$ will decrease further below $M_{SUSY}$, and the
difference between $M^{*}$ and $M_{SUSY}$ at the moment of the
transition can be significant, say one or two orders of
magnitude. Hence $M_{SUSY}$ can be $10^{16}\,GeV$ and this does
not create problems with the CMBR. Let us also notice the
oscillatory behavior of $H(t)$ after reaching the plateau, i.e.
when the system is about jumping into the FLRW phase.
Interestingly enough, this behavior could perhaps be related to
the reheating phenomenon, which is of course indispensable before
the universe stabilizes in the FLRW regime.

Overall, we arrive at a consistent picture of the graceful exit in
this inflationary scenario. No inflaton field is needed, but only
the dynamical work of gravity itself, provided one starts from a
renormalizable, fully conformal-invariant classical picture,
together with the trace anomaly of the matter fields at the
quantum level. Moreover, according to (\ref{f}), the obtained
picture is universal, for it does not depend on the choice of the
dilatation symmetry breaking scale $M$. If interpreted
physically, one can put constraints on $M$ using the macroscopic
forces mediated by the field $\sigma$, demanding that this forces
should have the sub-millimeter range, similarly as in
\cite{SS1}.\\

 \noindent {\bf Acknowledgments.} The work of J.S. has been
supported in part by MECYT and FEDER under project FPA2001-3598.
The work of I.Sh. has been supported by the research grant from
FAPEMIG and fellowship from CNPq.

%%%%%%%%%%%%%%%%%%%%%%%%%%%%%%%%%%%%%%%%%%%%%%%%%%%%%%%%%%%%

%\vskip 1mm

%\renewcommand{\baselinestretch}{1.1}
%\renewcommand{\baselinestretch}{2.5}

\end{document}